\documentclass[floats,aps,showpacs,twocolumn,superscriptaddress]{revtex4-1}

\usepackage[dvips]{graphicx}
\usepackage{amsfonts}
\usepackage{amsmath,amssymb}
\usepackage{dsfont}
\usepackage{placeins}
\usepackage{afterpage}
\usepackage{flafter} 
\usepackage{color} 
\usepackage{stmaryrd}

\newcommand{\A}{\mathcal{M}} 
\newcommand{\RCal}{\mathcal{R}} 
\newcommand{\TCal}{\mathcal{T}} 
\newcommand{\KCal}{\mathcal{K}} 
\newcommand{\LCal}{\mathcal{L}} 
\newcommand{\NCal}{\mathcal{N}} 
 
\newcommand{\q}{\boldsymbol{q}}
\newcommand{\p}{\boldsymbol{p}}
\newcommand{\qbar}{\boldsymbol{\bar{q}}} 
\newcommand{\pbar}{\boldsymbol{\bar{p}}}
\newcommand{\Q}{\boldsymbol{Q}}
\renewcommand{\P}{\boldsymbol{P}}

\newcommand{\J}{\mathbf{J}}
\newcommand{\bPhi}{\boldsymbol{\phi}}
\newcommand{\ba}{\mathbf{a}}
\newcommand{\bx}{\mathbf{x}}

\newcommand{\xbar}{\bar{x}}
\newcommand{\ybar}{\bar{y}}
\newcommand{\pxbar}{\bar{p}_x}
\newcommand{\pybar}{\bar{p}_y}

\newcommand{\bomega}{\boldsymbol{\omega}} % use boldsymbol to get the greek boldface
\newcommand{\K}{K}
\newcommand{\Ta}{T^{\ba}}
\newcommand{\h}{h} 

\newcommand{\w}{w}

\newcommand{\reg}{\text{reg}}
\newcommand{\Hreg}{H_\reg}
\newcommand{\HregNod}{H_\reg^0}
\newcommand{\HregNodCal}{\mathcal{H}_\reg}
\newcommand{\hCal}{\mathcal{F}} 
\newcommand{\HCal}{\mathcal{H}} 

\newcommand{\dif}{\text{d}}

\newcommand{\insertfigure}[4]{
  \begin{figure}[#1]
    \centering
    \includegraphics{#2}%
    \caption{#3}
    \label{#4}
  \end{figure}
}
\newcommand{\insertbigfigure}[4]{
  \begin{figure*}[#1]
    \centering
    \includegraphics{#2}%
    \caption{#3}
    \label{#4}
  \end{figure*}
}

\newcommand{\vrr}[2]{\left(\begin{array}{cc} #1 \\ #2 \end{array}\right)} 

\newcommand{\DD}[2]{\frac{\partial #1}{\partial #2}}

\begin{document}

\title{Integrable Approximation of Regular Islands: \\ The Iterative Canonical Transformation Method}

\author{Clemens L\"obner}
\affiliation{Technische Universit\"{a}t Dresden,
          Institut f\"{u}r Theoretische Physik and Center for Dynamics,
          01062 Dresden, Germany} 
\affiliation{Max-Planck-Institut f\"ur Physik komplexer Systeme, N\"othnitzer
Stra\ss{}e 38, 01187 Dresden, Germany}

\author{Steffen L\"ock}
\affiliation{Technische Universit\"{a}t Dresden,
          Institut f\"{u}r Theoretische Physik and Center for Dynamics,
          01062 Dresden, Germany} 
\affiliation{OncoRay - National Center for Radiation Research in Oncology, TU Dresden, 
  Fetscherstra\ss{}e 74, 01307 Dresden, Germany}

\author{Arnd B\"acker}
\affiliation{Technische Universit\"{a}t Dresden,
          Institut f\"{u}r Theoretische Physik and Center for Dynamics,
          01062 Dresden, Germany} 
\affiliation{Max-Planck-Institut f\"ur Physik komplexer Systeme, N\"othnitzer
Stra\ss{}e 38, 01187 Dresden, Germany}

\author{Roland Ketzmerick}       
\affiliation{Technische Universit\"{a}t Dresden,
          Institut f\"{u}r Theoretische Physik and Center for Dynamics,
          01062 Dresden, Germany} 
\affiliation{Max-Planck-Institut f\"ur Physik komplexer Systeme, N\"othnitzer
Stra\ss{}e 38, 01187 Dresden, Germany}

\date{\today}

\begin{abstract}

Generic Hamiltonian systems have a mixed phase space, 
where classically disjoint regions of regular and chaotic motion coexist.
We present an iterative method to construct an integrable approximation 
$\Hreg$, which resembles the regular dynamics of a given mixed system $H$ 
and extends it into the chaotic region. The method is based on the construction
of an integrable approximation in action representation which is then improved 
in phase space by iterative applications of canonical transformations. 
This method works for strongly perturbed systems and arbitrary degrees of freedom.
We apply it to the standard map and the 
cosine billiard.
\end{abstract}
\maketitle

\noindent

%%%%%%%%%%%%%%%%%%%%%%%%%%%%%%%%%%%%%%%%%%%%%%%%%%%%%%%%%%%%%%%%%%%%%%%%%%%%%%%

\section{Introduction}
\label{ch:Introduction}

\insertfigure{tb}{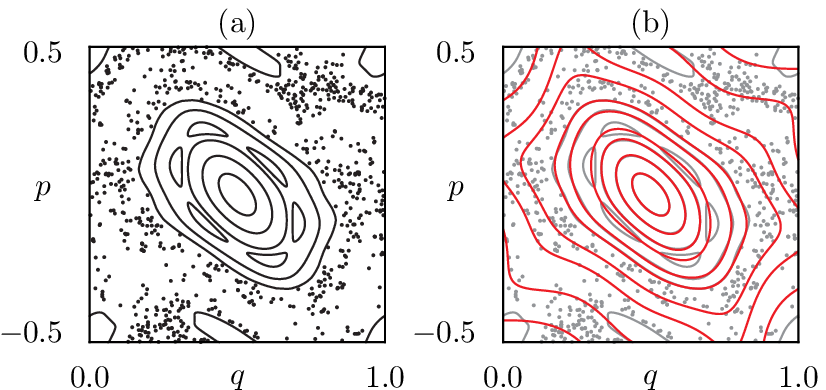}{(color online) 
(a) Phase space of the standard map, Eq.~\eqref{eq:StandardMap} at $K=1.25$, with regular orbits (black lines) and chaotic orbits (black dots)
and (b) of its integrable approximation $\Hreg$ (thin red lines) following from the iterative canonical transformation method.}{fig:IntroductionStandardMap} 

Classical Hamiltonian systems are an important class of dynamical systems with relevance in many areas of physics, e.g.\ celestial mechanics, accelerator physics, molecular physics, and semiclassical methods \cite{MeyHal1992,LicLie}.
A particular case are integrable systems \cite{ArnoldMechanics} where
the dynamics is restricted to invariant tori in phase space.
The other extreme 
is given by fully chaotic systems 
in which almost all trajectories show 
sensitive dependence on the initial conditions. 
Generically, however, Hamiltonian systems have a mixed phase 
space, in which regions of regular and chaotic motion coexist \cite{MarMey1974}.
This is illustrated in Fig.~\ref{fig:IntroductionStandardMap}(a) for a 2D symplectic map showing a \textit{regular island} composed of regular tori surrounded by the \textit{chaotic sea}. Within the regular island there is a rich self-similar substructure including nonlinear resonance chains and thin chaotic layers on all scales \cite{LicLie}.

For many applications it turns out to be extremely useful to replace the complicated fine details of the dynamics inside the regular island of a mixed Hamiltonian $H$ by an integrable approximation $\Hreg$.
This integrable system $\Hreg$ should resemble the dynamics 
in a regular island of $H$ as good as possible and smoothly interpolate through regions of nonlinear resonance chains and chaotic layers.
An example for such an integrable approximation is shown in 
Fig.~\ref{fig:IntroductionStandardMap}(b).
A number of methods have been established in the literature: 
For near-integrable systems one can use normal-form techniques \cite{Bir1927,Gus1966,Mey1974,SchWaaWig2006} or 
perturbative methods based on 
Lie transforms \cite{Dep1969,Car1981,LicLie,BroSchUll2002} or the Campbell-Baker-Hausdorff formula \cite{Sch1988}. 
For strongly perturbed 2D symplectic maps a method based on the frequency analysis and a series expansion for $\Hreg(q,p)$ 
was developed \cite{BaeKetLoe2010}.
For the approximation of individual tori by a Hamiltonian, see Refs.~\cite{ChaGarMil1976,KaaBinPRL1994}. 
If, however, one considers higher-dimensional systems with strongly perturbed regular islands, a new method for finding an overall $\Hreg$ is needed.

A current motivation for finding an integrable approximation is to describe the quantum tunneling from a regular island into the chaotic sea. 
This is possible using the fictitious integrable system approach \cite{BaeKetLoeSch2008,BaeKetLoe2010,LoeBaeKet2010,MerEtAl} 
which successfully predicts tunneling rates for 2D maps. 
An extension to higher-dimensional systems would require a suitable method for determining an integrable approximation $\Hreg$.
Other applications for integrable approximations are, e.g., in the field of toroidal magnetic devices \cite{HudDew1998}.

In this paper we introduce the \textit{iterative canonical transformation method} to construct an integrable approximation $\Hreg$ for a generic regular island. 
This method can be 
extended to arbitrary degrees of freedom. The resulting integrable approximation $\Hreg$ resembles 
the shape of the tori from the regular island and the dynamics on these tori. 
Furthermore, it interpolates through zones of nonlinear resonance chains and chaotic layers and extends the tori beyond the regular island, see Fig.~\ref{fig:IntroductionStandardMap}(b).
The method first finds an integrable approximation in action representation which mimics the frequency dependence within the regular island. Secondly, in phase space a sequence of canonical transformations is applied which optimizes the shape of and the dynamics on the tori.
To illustrate the method we apply it to a generic 2D symplectic map. An application to a higher-dimensional system is presented for a generic billiard. 

This paper is organized as follows: In Sec.~\ref{ch:GeneralMethod} we 
introduce the iterative canonical transformation method. 
We present its application to the standard map in 
Sec.~\ref{ch:StandardMap} and discuss its relation to other 
approaches. In Sec.~\ref{ch:CosineBilliard} 
we apply the approach to the generic cosine billiard.
A summary and outlook is given in Sec.~\ref{ch:Summary}. 

%%%%%%%%%%%%%%%%%%%%%%%%%%%%%%%%%%%%%%%%%%%%%%%%%%%%%%%%%%%%%%%%%%%%%%%%%%%%%%%
\section{Iterative canonical transformation method}
\label{ch:GeneralMethod}

Our aim is to approximate a given non-integrable Hamiltonian $H(\q,\p)$ with $f$ degrees of freedom by an integrable Hamiltonian $\Hreg(\q,\p)$. This integrable approximation $\Hreg$ should resemble the dynamics of $H$ in a regular island around a stable orbit, see Fig.~\ref{fig:IntroductionStandardMap}.

The construction of the integrable approximation $\Hreg(\q,\p)$ is performed in two steps. We first define an integrable approximation in action representation in Sec.~\ref{ch:GeneralActionRepresentation}. It is then transformed to a phase-space representation and is improved iteratively using canonical transformations in Sec.~\ref{ch:GeneralPhaseSpaceRepresentation}.

\subsection{Action representation}
\label{ch:GeneralActionRepresentation}

The time evolution of a Hamiltonian system in the $2f$-dimensional phase space $(\q,\p)$ is determined by Hamilton's equations 
\begin{align}
 \dot{\q} = \DD{H}{\p}(\q,\p),&\qquad \dot{\p} = -\DD{H}{\q}(\q,\p).\label{eq:HamiltonianMotion}
\end{align} For an integrable system with $f$ degrees of freedom one has $f$ constants of motion such that the dynamics takes place on $f$-dimensional tori. Likewise the regular motion of a non-integrable Hamiltonian $H$ is also confined to $f$-dimensional tori. To each torus the action $\J=(J_1,\hdots,J_f)$ with
\begin{align}
 J_i &:=\frac{1}{2\pi}\oint_{C_i} \p \,\dif \q \label{eq:GeneralActionIntegrals}
\end{align} can be associated. Here the $C_i$ are topologically independent closed paths on the torus. 

The starting point of the iterative canonical transformation method is the construction of an integrable approximation $\HregNodCal(\J)$ in action representation.
We obtain $\HregNodCal(\J)$ from the actions $\J$ of the tori of the original system $H$.
For a large number of regular tori $\tau$ with energy $E^\tau$ we determine a sufficiently long trajectory using Eq.~\eqref{eq:HamiltonianMotion} from which the action $\J^\tau$ can be determined numerically using Eq.~\eqref{eq:GeneralActionIntegrals}. For explicit examples see Secs.~\ref{ch:StandardMap} and \ref{ch:CosineBilliard}. Based on this data we express $\HregNodCal(\J)$ by a series expansion, e.g.\ a polynomial expansion, which minimizes 
\begin{align}
 \sum_\tau |E^\tau - \HregNodCal(\J^\tau)|^2.\label{eq:HregNodGeneralAnsatzEnergy}
\end{align} In order to smoothly interpolate through zones of nonlinear resonance chains, tori close to them have to be excluded. 

\subsection{Phase-space representation}
\label{ch:GeneralPhaseSpaceRepresentation}

In this section we first transform $\HregNodCal(\J)$ to the initial integrable approximation $\HregNod(\q,\p)$ in phase-space representation, such that it roughly approximates the regular tori of $H(\q,\p)$.
Then we search for a sequence of $N$ canonical transformations
\begin{align}
 T_n: (\q,\p) &\mapsto (\q',\p'),\quad n=1,\hdots,N\label{eq:GeneralTOne}
\end{align} which in each step gives
\begin{align}
 \Hreg^{n}(\q,\p) &= \Hreg^{0}[T_1^{-1}\circ\hdots\circ T_n ^{-1}(\q,\p)],\label{eq:GeneralHreg1ByPluggingInTransformation}
\end{align} such that the tori of $\Hreg^n$ show a better agreement with the corresponding tori of $H$. 
As the transformations $T_n$ preserve integrability we obtain a sequence of integrable approximations $\Hreg^n$. To find such a transformation we introduce a family of canonical transformations which can be varied smoothly by a parameter. We define a cost function which measures the quality of the transformation. The optimal transformation can then be determined by minimizing the cost function. 

\subsubsection{Initial integrable approximation}
\label{ch:GeneralInitialIntegrableApproximation}

Starting from $\HregNodCal(\J)$ we choose a simple canonical transformation
\begin{align}
 T_0 : (\bPhi,\J) &\mapsto (\q,\p),\label{eq:GeneralTransformationToqp} 
\end{align} which maps the tori of $\HregNodCal(\J)$ to the neighborhood of the tori of $H(\q,\p)$ with the same action $\J$ such that they have roughly the same shape. This canonical transformation leads to the \textit{initial integrable approximation}
\begin{align}
 \HregNod(\q,\p) &= \HregNodCal[ \J(\q,\p)] .\label{eq:GeneralHamiltonianTransformationToqp}
\end{align} It can be determined, e.g., from the linearized dynamics of $H(\q,\p)$ at the center of the island. 
Explicit examples are given in Secs.~\ref{ch:StandardMap} and \ref{ch:CosineBilliard}.

\subsubsection{Family of canonical transformations}
\label{ch:GeneralImprovementFamilyOfCanonicalTransformations}

We define a family of canonical transformations $\Ta$ depending on a parameter vector $\ba=(a_1,\hdots ,a_r)\in\mathbb{R}^r$. This is realized using a type $2$ generating function of the form
\begin{align}\label{eq:GeneralGenerator}
 F^\ba(\q,\p') &= \sum_{i=1}^f q_ip_i' + \sum_{\nu=1}^r a_\nu G_\nu (\q,\p').
\end{align} The concrete family of transformations is defined by choosing a set of $r$ independent functions $G_\nu$. This choice depends on the considered problem, see Secs.~\ref{ch:StandardMap} and \ref{ch:CosineBilliard} for examples. The canonical transformation $\Ta$ is then defined by the set of equations 
\begin{align}
  \q'& = \DD{F^\ba}{\p'}(\q,\p') = \q + \sum_{\nu=1}^r a_\nu\DD{G_\nu}{\p'}(\q,\p'), \label{eq:Type21}\\
  \p & = \DD{F^\ba}{\q}(\q,\p') = \p' + \sum_{\nu=1}^r a_\nu\DD{G_\nu}{\q}(\q,\p'), \label{eq:Type22}
\end{align} which implicitly connect the old variables $(\q,\p)$ to the new variables $(\q',\p')$ \cite{GoldsteinMechanics}. For $\ba=\mathbf{0}$ one obtains the identity transformation. As the tori of $\HregNod$ and $H$ roughly agree, we consider near-identity transformations with 
\begin{align}
 |\ba|  &\ll 1 \label{eq:SmallParameters}
\end{align} only. This ensures, according to the implicit function theorem, the invertibility of Eqs.~\eqref{eq:Type21} and \eqref{eq:Type22} such that the transformation $\Ta$ is well-defined in a sufficiently large phase-space region. 

\subsubsection{Iterative Improvement}
\label{ch:GeneralEstimatorFunction}

We now iteratively improve the initial integrable approximation $\HregNod$ from Sec.~\ref{ch:GeneralInitialIntegrableApproximation}. For a given integrable approximation $\Hreg^n$, Eq.~\eqref{eq:GeneralHreg1ByPluggingInTransformation}, we determine a transformation $T_{n+1}$ leading to a better integrable approximation $\Hreg^{n+1}$. 
We introduce a cost function $L(\ba)$ in the parameter space of $\ba$, which measures how well a particular transformation $T_{n+1}=\Ta$ improves the current integrable approximation $\Hreg^n$,
\newcommand{\xttau}{\bx_t^{\tau}}
\newcommand{\xttaun}{\bx_t^{\tau,n}}
\newcommand{\xttauN}{\bx_t^{\tau,N}}
\newcommand{\xttaunminus}{\bx_t^{\tau,n-1}}
\newcommand{\xttaunod}{\bx_t^{\tau,0}}
\newcommand{\qttau}{\q_t^{\tau}}
\newcommand{\pttau}{\p_t^{\tau}}
\newcommand{\qttaun}{\q_t^{\tau,n}}
\newcommand{\pttaun}{\p_t^{\tau,n}}
\newcommand{\qttaureg}{\q_t^{\tau,\reg}}
\newcommand{\pttaureg}{\p_t^{\tau,\reg}}
\begin{align}
 L(\ba) &= \frac{1}{\NCal}\sum_\tau \sum_t \left[\xttau - \Ta (\xttaun)\right]^2.\label{eq:GeneralCostFunction}
\end{align} Here we sum over a suitable set of tori $\tau$ of $H$ which is not necessarily the same set as used to minimize Eq.~\eqref{eq:HregNodGeneralAnsatzEnergy}. 
For each torus $\tau$, we sum over a set of its points from a solution $\xttau=(\q^\tau _t,\p^\tau _t)$ of $H$, Eq.~\eqref{eq:HamiltonianMotion}, at discrete times $t$ which gives $\NCal$ points in total. For each time $t$ we compute the distance of this solution $\xttau$ to the transformation $\Ta (\xttaun)$ of the corresponding solution $\xttaun$ of $\Hreg^n$ from the previous iteration step, 
\begin{align}
 \xttaun &= T_n (\xttaunminus), \qquad n\geq 1,\label{eq:ReferencePointsRecoursion}
\end{align} which lies on the integrable torus with the same action $\J^\tau$ as $\tau$. We evaluate the transformation $T_n$ obtained in the previous iteration step numerically using Eqs.~\eqref{eq:Type21} and \eqref{eq:Type22}. For the first iteration step we use
\begin{align}
 \xttaunod &= T_0 (\bPhi^\tau_0 + \bomega^\tau t, \J^\tau), \qquad n=0,\label{eq:ReferencePointsInitial}
\end{align} with $T_0$ given by Eq.~\eqref{eq:GeneralTransformationToqp}. The initial angle $\bPhi^\tau_0$ is determined, such that the initial point $\bx ^{\tau,0}_0$ is closest to $\bx_0 ^\tau$. Note that in Eq.~\eqref{eq:ReferencePointsRecoursion} we use the numerically determined frequency $\bomega^\tau$ of $H$ on the torus $\tau$, instead of the approximated frequency $\bomega=\partial \HregNodCal/\partial \J$.

We determine an approximation of the cost function $L(\ba)$, Eq.~\eqref{eq:GeneralCostFunction}, with the help of a first order approximation of the canonical transformation, Eqs.~\eqref{eq:Type21} and \eqref{eq:Type22}, 
\begin{align}
 \q'(\q,\p) &= \q+\sum_{\nu=1}^r a_\nu \DD{G_\nu}{\p}(\q,\p) +O(\ba^2), \label{eq:GeneralTransformationSecondOrder1}\\
 \p'(\q,\p) &= \p-\sum_{\nu=1}^r a_\nu \DD{G_\nu}{\q}(\q,\p) +O(\ba^2).\label{eq:GeneralTransformationSecondOrder2}
\end{align} This leads to the approximation
\begin{align}
 L(\ba) &\approx L(\mathbf{0}) -\frac{2}{\NCal}\sum_{\nu=1}^r B_\nu a_\nu +\frac{1}{\NCal}\sum_{\mu,\nu=1}^ra_\mu C_{\mu\nu}a_\nu,\label{eq:GeneralEstimatorSecondOrderApproximation}
\end{align} where
\newcommand{\dtau}{d_\tau ^0} % shorthand for these eqs only  
\begin{align}
 B_\nu &= \sum_{\tau,t}(\qttau-\qttaun)\DD{G_\nu}{\p}(\qttaun,\pttaun)\label{eq:CoeffBnu}\\
 &-\sum_{\tau,t}(\pttau-\pttaun)\DD{G_\nu}{\q}(\qttaun,\pttaun), \nonumber\\
 C_{\mu\nu} &= \sum_{\tau,t} \DD{G_\mu}{\p}(\qttaun,\pttaun)\DD{G_\nu}{\p}(\qttaun,\pttaun) \label{eq:CoeffCmunu}\\
 &+\sum_{\tau,t} \DD{G_\mu}{\q}(\qttaun,\pttaun)\DD{G_\nu}{\q}(\qttaun,\pttaun), \nonumber
\end{align} which depend on the chosen tori $\tau$ and the sample points $\xttau$ and $\xttaun$. 

The optimal parameter $\ba$ is the solution of the extremal condition $\partial L/\partial a_\mu =0$, which using the approximation~\eqref{eq:GeneralEstimatorSecondOrderApproximation} becomes a system of linear equations
\begin{align}
 \sum_{\nu=1}^r C_{\mu\nu}a_\nu &= B_\mu,\qquad \mu=1,\hdots,r,\label{eq:GeneralLinearSystemOfEquations}
\end{align} that can be solved for $\ba$ by matrix inversion. This solution defines the new canonical transformation $\Ta$. 

If the resulting transformation $\Ta$ is not invertible in the relevant phase-space region, then Eq.~\eqref{eq:SmallParameters} suggests to scale down the solution parameters according to
\begin{align}
 \ba &\mapsto \eta\ba,\label{eq:Damping}
\end{align} using a sufficiently small damping factor $\eta\in]0,1[$. 
In this way the damping allows for finding large transformations built up from small steps. 
The required number $N$ of iteration steps will increase roughly by a factor of $1/\eta$.

In the $n$-th step the resulting transformation $T_{n+1}:=\Ta$ leads to a better approximation $\Hreg^{n+1}$ of $H$, Eq.~\eqref{eq:GeneralHreg1ByPluggingInTransformation}. Typically after a finite number $N$ of iterations the cost function saturates and one can stop the iterative process. This leads to an optimized integrable approximation $\Hreg^N$ of $H$. Note that $\Hreg^N$ is not given in an analytic form, as the transformations $T_n^{-1}$ in Eq.~\eqref{eq:GeneralHreg1ByPluggingInTransformation} have to be evaluated numerically from Eqs.~\eqref{eq:Type21} and \eqref{eq:Type22}. 
The necessity of an iterative approach follows from two reasons. 
First, the exact minimization problem, Eq.~\eqref{eq:GeneralCostFunction}, is replaced by an approximate one, Eq.~\eqref{eq:GeneralEstimatorSecondOrderApproximation}. Secondly, the transformations $\Ta$ typically do not form a group, i.e. $\Ta \circ T^{\ba'}\neq T^{\ba+\ba'}$. Hence the application of multiple steps provides new solutions which cannot be obtained by using just one step. This is explicitly employed when using the damping factor. 
An algorithmic overview  of this iterative canonical transformation method is given in App.~\ref{app:AlgorithmicOverview}.

\section{Application to 2D maps}
\label{ch:StandardMap} 

In this chapter we apply the iterative canonical transformation method to the simplest class of non-integrable systems, given by 2D symplectic maps. The paradigmatic model of such a map $M$ with a mixed phase space is the standard map \cite{Chi1979} which is defined by
\begin{align}
 \vrr{q'}{p'} &= \vrr{q +p}{p +\frac{\K}{2\pi}\sin (2\pi q')}, \label{eq:StandardMap}
\end{align} on the phase space $(q,p)\in [0,1[ \times [-0.5,0.5[$ with periodic boundary conditions. For $0<K<4$ one has a stable fixed point at $(q^*,p^*)=(\tfrac{1}{2},0)$. 
We consider the standard map for the parameter $\K=1.25$. The phase space is shown in Fig.~\ref{fig:IntroductionStandardMap}(a). Around the fixed point $(q^*,p^*)$ one finds an island of regular tori which is embedded in a chaotic sea.

We now determine an integrable approximation $\Hreg(q,p)$ that mimics the dynamics in the regular island of the standard map. We follow the steps of the iterative canonical transformation method as described in Sec.~\ref{ch:GeneralMethod}.

\subsection{Action representation}
\label{ch:StandardMapActionRepresentation}

We first determine the integrable approximation $\HregNodCal(J)$ in action representation. We cannot use Eq.~\eqref{eq:HregNodGeneralAnsatzEnergy} for that purpose as for maps the energy is not defined. Instead we use the frequency function $\omega(J)$ which is uniquely related to $\HregNodCal(J)$ according to
\begin{align}
 \omega(J) &= \DD{\HregNodCal}{J}(J)\label{eq:HregNodCalOmegaIntegral}
\end{align} and minimize
\begin{align}
 \sum_\tau \left| \omega_\tau - \omega(J_\tau)\right|^2. \label{eq:HregNodGeneralAnsatzFrequency}
\end{align} Here we compute the frequencies $\omega_\tau$ and the actions $J_\tau$ from the 
regular tori $\tau$ of the map $M$ by choosing a set of initial conditions on a line from the center of the regular island to its border (the outermost torus shown in Fig.~\ref{fig:IntroductionStandardMap}(a)). Specifically, we use the points $(q^*+\tfrac{\tau}{60}\Delta q,p^*)$ with $\tau=1,\hdots,60$ and $\Delta q = 0.293$. For each initial condition we calculate $10^4$ iterates. From these points we determine the frequency $\omega_\tau$ \cite{Las1992,BarBazGio1995} and the action $J_\tau$, Eq.~\eqref{eq:GeneralActionIntegrals}, see the black dots in Fig.~\ref{fig:StandardMapNumerics}. As mentioned in Sec.~\ref{ch:GeneralActionRepresentation} data points near nonlinear resonances are omitted. Using Eq.~\eqref{eq:HregNodGeneralAnsatzFrequency} we fit a polynomial 
\begin{align}
 \omega(J) &= \sum_{k=0}^\mathcal{K} \alpha_kJ^k\label{eq:StandardMapPolynomial}
\end{align} to the remaining data. Up to order $\mathcal{K}=5$ (red curve in Fig.~\ref{fig:StandardMapNumerics}) we find a significant improvement of the fit. For too high orders $\mathcal{K}$ the extrapolation of $\omega(J)$ to actions beyond the border of the regular island strongly depends on $\mathcal{K}$. Finally, we integrate Eq.~\eqref{eq:HregNodCalOmegaIntegral} to obtain an integrable Hamiltonian 
\begin{align}
 \HregNodCal(J) &= \sum_{k=0}^\mathcal{K} \frac{\alpha_k}{k+1}J^{k+1}.
\end{align}
\insertfigure{b}{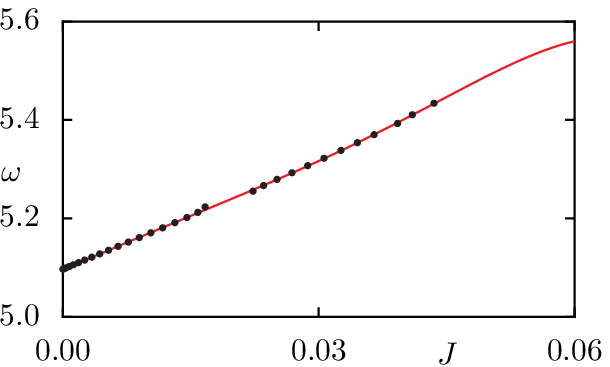}{(color online) Numerically determined frequency $\omega$ vs. action $J$ (dots) fitted by Eq.~\eqref{eq:StandardMapPolynomial} with $\mathcal{K}=5$ (red line) and extrapolated beyond the border of the regular island.}{fig:StandardMapNumerics} 

\subsection{Phase-space representation}
\label{ch:StandardMapPhaseSpaceRepresentation}
Following Sec.~\ref{ch:GeneralPhaseSpaceRepresentation} we now determine an initial integrable approximation $\HregNod(q,p)$ in phase-space representation, which we then improve by an iterative application of canonical transformations.

\subsubsection{Initial integrable approximation}
\insertbigfigure{bt}{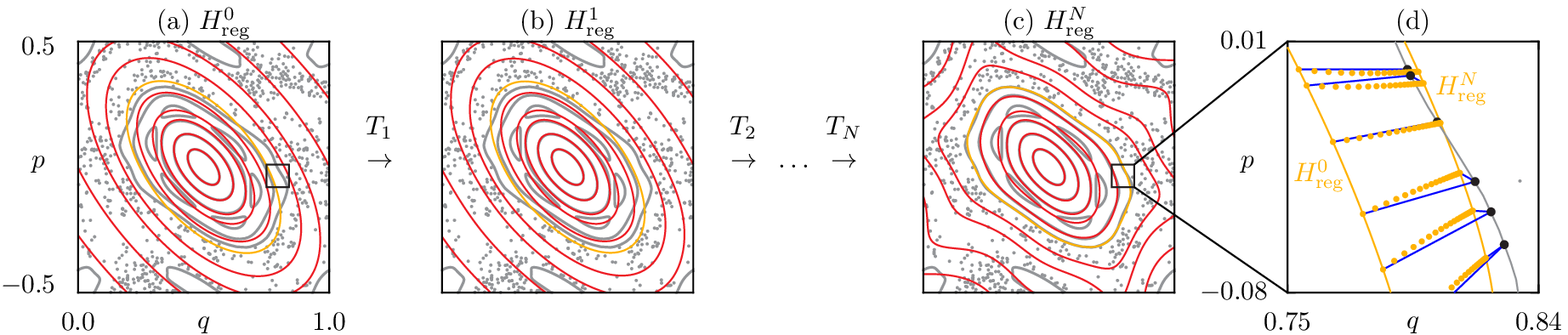}{(color online) The phase space of the standard map, Eq.~\eqref{eq:StandardMap}, at $\K=1.25$ (thick gray lines and dots) compared to the tori (thin colored lines) of (a) the initial integrable approximation $\HregNod$, (b) its transformation after the first iteration step $T_1$, and (c) after the final iteration step $T_N$, $N=60$. (d) Magnification of the border torus of $\HregNod$ and $\Hreg^N$ (thin green lines) and the standard map (thick gray line) together with the individual points $\xttaun$ (small green dots) of $\Hreg^n$ approaching the reference points $\xttau$ (big black dots) of the standard map. Straight lines (blue) indicate the initial and final distances that contribute to the cost function, Eq.~\eqref{eq:GeneralCostFunction}.}{fig:StandardMapTransformationCascade}
To obtain $\HregNod(q,p)$, we transform the dynamics of $\HregNodCal(J)$ to new coordinates $(q,p)$, according to the linearized dynamics of $M$ around the fixed point $(q^*,p^*)$. This is realized by the transformation
\begin{align}
 T_0:\vrr{\phi}{J}\mapsto\vrr{q}{p} &= \vrr{q^*}{p^*} + \RCal \vrr{\sqrt{2J}\cos\phi}{\sqrt{2J}\sin\phi}, \label{eq:StandardMapTransformationToQPForHregNod}
\end{align} where
\begin{align} 
 \RCal &:= \vrr{\cos\theta & -\sin\theta}{\sin\theta & \cos\theta} \vrr{1/\sqrt{\sigma} & 0}{0 & \sqrt{\sigma}}.\label{eq:StandardMapRCal}
\end{align}
This transformation generates elliptic tori with a tilting angle $\theta$ and the  axial ratio $\sigma$ as shown by the thin lines in Fig.~\ref{fig:StandardMapTransformationCascade}(a). The parameters $\theta$ and $\sigma$ are properties of the linearized system and are given by \cite{LicLie}
\begin{align}
 \tan2\theta &= \frac{\A_{11}-\A_{22}}{\A_{12}+\A_{21}},\\
 \sigma^2 &= \frac{|\A_{12}-\A_{21}|-c}{|\A_{12}-\A_{21}|+c},\label{eq:SigmaFromMonodromyMatrix}
\end{align} with
\begin{align}
  c &= \sqrt{(\A_{12}+\A_{21})^2+(\A_{22}-\A_{11})^2}.\label{eq:SigmaConstantc}
\end{align} Here $\A$ is the monodromy matrix of the standard map at the stable fixed point $(q^*,p^*)$, 
\begin{align} 
 \A &= \left.\vrr{\DD{q'}{q} &  \DD{q'}{p} }{\DD{p'}{q} & \DD{p'}{p}}\right|_{(q^*,p^*)}=\vrr{1 & 1}{-\K & 1-\K}. 
\end{align}

The inverse of the canonical transformation~\eqref{eq:StandardMapTransformationToQPForHregNod} gives $J(q,p)$ and generates a Hamiltonian $\HregNod(q,p)$ according to Eq.~\eqref{eq:GeneralHamiltonianTransformationToqp}. We stress that in contrast to the linearized dynamics of $M$, $\HregNod(q,p)$ contains the global frequency information of the regular island. Note that $\HregNod$ does not obey the periodic boundary conditions of the map, which is not relevant for approximating the regular island. The comparison in Fig.~\ref{fig:StandardMapTransformationCascade}(a) shows that the tori of the initial integrable approximation $\HregNod(q,p)$ agree with the tori of $M$ in the vicinity of the fixed point only. In the remaining part of the island $\HregNod$ needs to be improved.

\subsubsection{Family of canonical transformations}
First we define a family of canonical transformations $\Ta$ by choosing a generator basis $G_\nu$ to be used in Eq.~\eqref{eq:GeneralGenerator}. Since the tori of $M$ and $\HregNod$ are symmetric with respect to the fixed point, the considered transformations should conserve this symmetry and commute with the symmetry operation
\begin{align}
 \quad \vrr{q-q^*}{p-p^*} &\mapsto \vrr{-(q-q^*)}{-(p-p^*)},
\end{align} i.e. we restrict to generators which satisfy
\begin{align}
  G_\nu(-(q-q^*),-(p'-p^*)) &= G_\nu(q-q^*,p'-p^*).\label{eq:StandardMapSymmetryGenerator}
\end{align} We choose the Fourier ansatz
\begin{align}
  F^\ba(q,p') &= qp' \label{eq:StandardMapGeneratingFunction}\\
             & + \sum_{\nu_1=0}^{\NCal_q}\sum_{\nu_2=0}^{\NCal_p}a^+_{\nu_1 \nu_2}f^+_{\nu_1}\left(\frac{q-q^*}{\LCal_q}\right)f^+_{\nu_2}\left(\frac{p'-p^*}{\LCal_p}\right)\nonumber\\
             & + \sum_{\nu_1=1}^{\NCal_q}\sum_{\nu_2=1}^{\NCal_p}a^-_{\nu_1 \nu_2}f^-_{\nu_1}\left(\frac{q-q^*}{\LCal_q}\right)f^-_{\nu_2}\left(\frac{p'-p^*}{\LCal_p}\right),\nonumber
\end{align} with basis functions
\begin{align}
 f^+_\nu(x) &= \cos\left(2\pi\nu x\right),\label{eq:FourierBasisPlus}\\
 f^-_\nu(x) &= \sin\left(2\pi\nu x\right).\label{eq:FourierBasisMinus}
\end{align} Thus the coefficients to be optimized are $\ba=(a^+ _{\nu_1,\nu_2},a^- _{\nu_1,\nu_2})$ with $a^+ _{0 0}=0$. The orders $\NCal_{q,p}$ and the periods $\LCal_{q,p}$ of the basis functions can still be chosen.

\subsubsection{Iterative Improvement} 
We now perform the iterative improvement in order to transform the tori of $\HregNod$ closer to the tori of $M$. First we compute the coefficients $B_\nu$ and $C_{\mu\nu}$ of the cost function, Eqs.~\eqref{eq:CoeffBnu} and \eqref{eq:CoeffCmunu}, summing over all tori determined in Sec.~\ref{ch:StandardMapActionRepresentation} and over times $t=0,1,2,\hdots,10^4$. For the generating function~\eqref{eq:StandardMapGeneratingFunction} we choose period lengths $\LCal_{q,p}\approx 1$ and low orders $\NCal_{q,p}$, specifically $\LCal_q=\LCal_p=1.33$ and $\NCal_q=\NCal_p=2$. Finally we obtain a solution $\ba_1$ of Eq.~\eqref{eq:GeneralLinearSystemOfEquations} which we rescale using the strong damping factor $\eta=0.05$, Eq.~\eqref{eq:Damping}. This solution $\ba_1$ defines the first canonical transformation $T_1 = T^{\ba_1}$. As shown in Fig.~\ref{fig:StandardMapTransformationCascade}(b) this transformation slightly deforms the tori of the initial integrable approximation $\HregNod$. 

\insertfigure{tb}{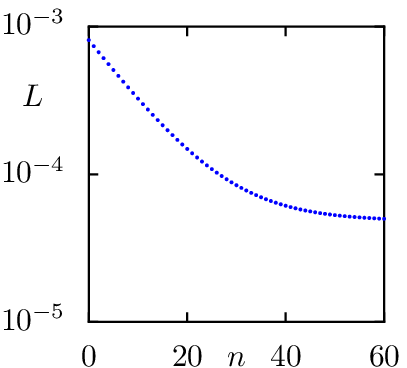}{(color online) 
Cost function $L$ vs. iteration step $n$. }{fig:StandardMapCostFunction} 

We repeat this procedure iteratively to obtain a sequence of transformations $(T_1,T_2,\hdots)$ leading to improved integrable approximations $(\Hreg^1,\Hreg^2,\hdots)$, see Fig.~\ref{fig:StandardMapTransformationCascade}. To quantify this iterative improvement we evaluate the cost function $L$ after each iteration step $n$, see Fig.~\ref{fig:StandardMapCostFunction}. For large $n$ the cost function saturates as $T_n$ converges to the identity transformation and the iteration is stopped after $N=60$ steps. The final integrable approximation $\Hreg^N (q,p)$ closely resembles the dynamics of the original map, as shown in Fig.~\ref{fig:StandardMapTransformationCascade}(c) for the shape of the tori and in Fig.~\ref{fig:StandardMapTransformationCascade}(d) for the individual points that are used for the cost function, Eq.~\eqref{eq:GeneralCostFunction}.

One would expect further optimization of the results by choosing higher orders $\NCal_{q,p}$ in Eq.~\eqref{eq:StandardMapGeneratingFunction} which increases the number of parameters of the transformation $\Ta$. However, it turns out that increasing $\NCal_{q,p}$ requires very small damping factors $\eta$ and hence more iteration steps, which reduces the performance. A possible improvement might be the increase of $\NCal_{q,p}$ only during the last iteration steps.

\subsection{Comparison to other methods}

\insertfigure{tb}{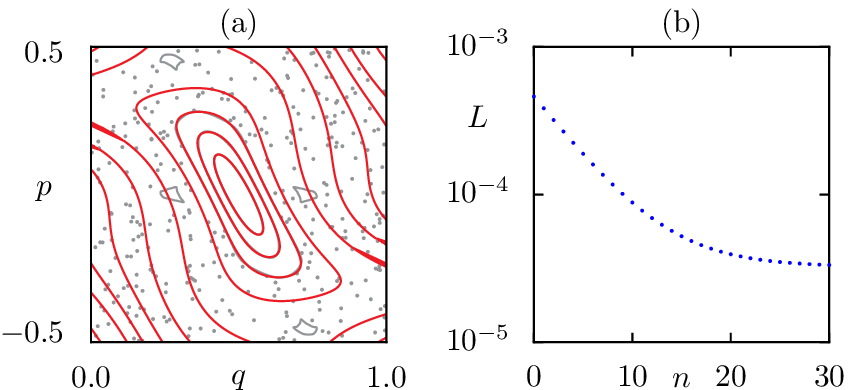}{(color online) (a) Phase space of the standard map, Eq.~\eqref{eq:StandardMap}, at $\K=2.9$ (thick gray lines and dots) and its integrable approximation (thin red lines) obtained using $N=30$ steps with the parameters $\LCal_q=2.86$, $\LCal_p=1.33$, $\NCal_q=1$, $\NCal_p=2$ in Eq.~\eqref{eq:StandardMapGeneratingFunction} and a damping factor $\eta=0.1$. (b) Cost function $L$ vs. iteration step $n$.}{fig:StandardMapOverlayAndDistanceK29} 

We now compare the iterative canonical transformation method with two other approaches.
A well known method for approximating Hamiltonian systems is the perturbation theory by means of Lie transforms which was introduced by Deprit \cite{Dep1969}. Tutorials can be found in Refs.~\cite{Car1981,LicLie} and in particular in Ref.~\cite{BroSchUll2002} the case of 2D maps is treated. 
One finds that this method fails to approximate the regular island of the standard map at $\K=2.9$ due to the strong perturbation \cite[Fig.~2(b)]{BaeKetLoe2010}, which is no obstacle for our method, as we show in Fig.~\ref{fig:StandardMapOverlayAndDistanceK29}.

For 2D maps a different method was introduced \cite{BaeKetLoe2010} which is also based on the frequency analysis of the main regular island. It gives results which are of the same quality as those presented in this chapter, \cite[Fig.~3(a)]{BaeKetLoe2010}.
However, this method cannot be extended to higher dimensions. 
In contrast the iterative canonical transformation method can be applied to systems of arbitrary dimension, which we demonstrate for the case of the 2D cosine billiard in the next section. 

%%%%%%%%%%%%%%%%%%%%%%%%%%%%%%%%%%%%%%%%%%%%%%%%%%%%%%%%%%%%%%%%%%%%%%%%%%%%%%%

\section{Application to billiards}
\label{ch:CosineBilliard}

We now apply the iterative canonical transformation method, introduced in Sec.~\ref{ch:GeneralMethod}, to find an integrable approximation for two-dimensional billiards. So far, integrable approximations of billiards have been constructed for near-integrable cases \cite{Cre2007} and for special geometries, in particular for the mushroom and the annular billiard \cite{BaeKetLoe2010, BaeKetLoe2008mush}, where the integrable approximation is the circular billiard. Here we demonstrate the applicability of our approach to generic billiards.

Two dimensional billiards are given by a point particle of mass $M$ moving along straight lines inside a domain $\Omega\subset\mathbb{R}^2$ with elastic reflections at the boundary. This motion is described by the Hamiltonian
\begin{align} 
 H(\q,\p) &= \left\{ \begin{array}{cl}\displaystyle{\frac{\p^2}{2M}} & \quad \q\in\Omega \\ \infty & \quad \q\notin\Omega\end{array} \right. , \label{eq:BilliardHamiltonian}
\end{align} where we set $2M=1$ in the following.

Billiards belong to the class of Hamiltonians with the scaling property  
\begin{align}
 H(\q,\lambda\p) &= \lambda^2H(\q,\p). \label{eq:ScalingHamiltonian}
\end{align} 
In such systems one can relate any solution of energy $E$ to a corresponding solution of energy $E=1$, see App.~\ref{app:Scaling}. Therefore we need information of trajectories from the billiard at energy $E=1$ only.
In the following we require that the integrable approximation $\Hreg(\q,\p)$ also belongs to the class of scaling Hamiltonians, Eq.~\eqref{eq:ScalingHamiltonian}.

\insertfigure{tb}{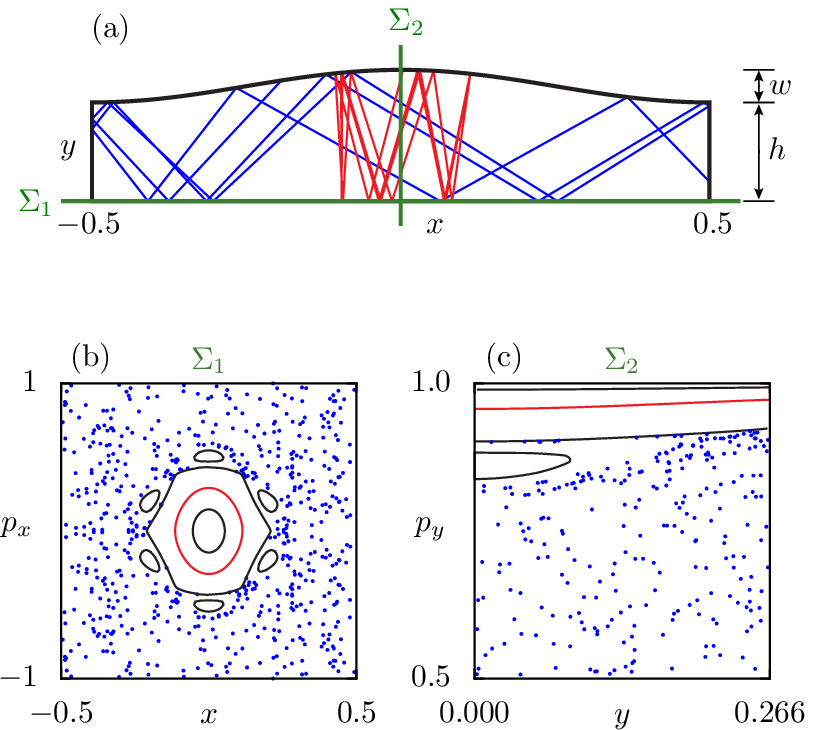}{(color online) (a) Cosine billiard in position space for $\h=0.2$ and $\w=0.066$ with a regular trajectory (dark red line), a chaotic trajectory (dashed blue line), and the positions of the Poincar\'e sections $\Sigma_1$ and $\Sigma_2$ (light green lines). (b) Dynamics on $\Sigma_1$ with a chaotic orbit (blue dots) and regular orbits (lines). (c) Same as (b) for $\Sigma_2$, zoomed to $p_y\in[0.5,1]$.}{fig:CosineBilliardIntroduction}
As a generic example we consider the cosine billiard \cite{StiTheses,LunKroRodHer1996,BaeSchSti1997,BaeManHucKet2002}, see Fig.~\ref{fig:CosineBilliardIntroduction}(a), which is given by a rectangle whose upper boundary is replaced by the curve
\begin{align}
 y &= r(x) := \h+\frac{\w}{2}\left[1+\cos(2\pi x)\right] \label{eq:CosineBilliardBoundary}
\end{align}  with $x\in [-0.5, 0.5]$. As parameters we use $\h=0.2$ and $\w=0.066$ leading to a generic mixed phase space. In order to visualize the dynamics which takes place on the energy shell $E=1$ in the 4D phase space we introduce two Poincar\'e sections $\Sigma_1$ $(y=0,p_y>0)$ and $\Sigma_2$ $(x=0,p_x>0)$ indicated by the light green lines in Fig.~\ref{fig:CosineBilliardIntroduction}(a). 
Figures~\ref{fig:CosineBilliardIntroduction}(b) and (c) show the intersections of some trajectories with the Poincar\'e sections $\Sigma_1$ and $\Sigma_2$, respectively. In each of these sections the chaotic orbit fills a two-dimensional region, whereas the regular orbits form one-dimensional sets. This indicates the structure of the regular tori, which are two-dimensional manifolds in the 4D phase space. 

In the following we determine an integrable approximation $\Hreg$ for the cosine billiard using the iterative canonical transformation method introduced in Sec.~\ref{ch:GeneralMethod}.

\subsection{Action representation}
\label{ch:CosineBilliardActionRepresentation}

To define the integrable approximation $\HregNodCal(\J)$ in action representation, we first compute points on a set of tori of $H$. As these tori are 2D manifolds in the 4D phase space, there exist 2 independent directions perpendicular to a given torus. Due to the restriction $E=1$ just one direction remains. Thus the relevant part of phase space can be explored using initial conditions along a 1D curve of constant energy. For this curve we choose a fixed position at the top of the cosine billiard and a momentum parametrized by an angle $\vartheta$,
\begin{align}
 x_0 &= 0, \\
 y_0 &= \h+\w, \\
 p_{x0} &=\sin\vartheta,\\
 p_{y0} &= -\cos\vartheta, 
\end{align} with  $\vartheta \in ]0,0.3926]$. For 100 initial conditions on this curve we calculate a trajectory with $10^4$ reflections, sampling the tori $\tau$. We compute their intersections with $\Sigma_1$ and $\Sigma_2$ to determine the actions $\J^\tau=(J_1^\tau,J_2^\tau)$ using Eq.~\eqref{eq:GeneralActionIntegrals}, see the dots in Fig.~\ref{fig:CosineBilliardActionsAndFrequencies}(a). For an alternative approach see Ref.~\cite{VebRobLiu1999}. 
We compute the frequencies $\bomega^\tau=(\omega_1^\tau,\omega_2^\tau)$ for each torus $\tau$ \cite{Las1992,BarBazGio1995}. 

\insertfigure{tb}{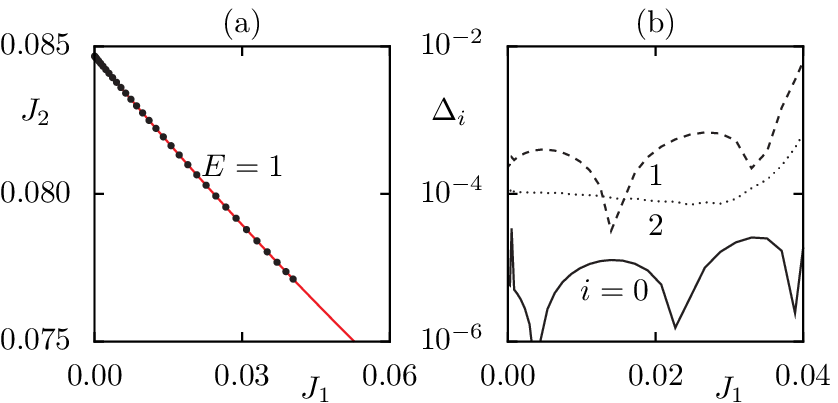}{(color online) (a) Numerically determined actions $\J^\tau=(J^\tau_1,J^\tau_2)$ of the cosine billiard (dots) and contour line $E=1$ (red) of the integrable Hamiltonian $\HregNodCal(\J)$, Eq.~\eqref{eq:HregNodScalingForm}. (b) Relative error $\Delta_0$ of the energy (solid) and the frequencies $\Delta_1$ (dashed) and $\Delta_2$ (dotted).}{fig:CosineBilliardActionsAndFrequencies}
According to Sec.~\ref{ch:GeneralActionRepresentation} we now determine an integrable Hamiltonian $\HregNodCal(\J)$, which connects the actions to the corresponding energies. The ansatz for this integrable Hamiltonian has to be chosen in accordance with the scaling property~\eqref{eq:ScalingHamiltonian} of $H$, which implies (see App.~\ref{app:Scaling}) 
\begin{align}
 \HregNodCal(\lambda \J) &= \lambda^2 \HregNodCal(\J).\label{eq:ScalingHamiltonianHregNodCal}
\end{align} Therefore it can be written, e.g., as
\begin{align}
 \HregNodCal(\J) &= J_2 ^2 \cdot \hCal(J_1/J_2)\label{eq:HregNodScalingForm}
\end{align} with some function $\hCal$. We express $\hCal$ as a power series
\begin{align}
 \hCal(J_1/J_2) &= \sum_{k=0}^\KCal \alpha_k \cdot (J_1/J_2)^k.
\end{align} This ansatz with $k\geq 0$ ensures finite energies in the center of the regular island, where $J_1=0$, see Fig.~\ref{fig:CosineBilliardIntroduction}(b). We determine the coefficients $\alpha_k$ such that Eq.~\eqref{eq:HregNodGeneralAnsatzEnergy} is minimized. For $\KCal=2$ we show in Fig.~\ref{fig:CosineBilliardActionsAndFrequencies}(a) the contour line $E=1$ of $\HregNodCal(\J)$ in good agreement with the numerical actions $\J^\tau$. In Fig.~\ref{fig:CosineBilliardActionsAndFrequencies}(b) we plot the relative errors of the energy $\Delta_0=|\left[\HregNodCal(\J^\tau)-E\right]/E|$ for $E=1$ and of the frequencies $\Delta_i=|\left[\partial\HregNodCal(\J^\tau)/\partial J_i - \omega^\tau_i\right]/\omega^\tau_i |$ with $i=1,2$. Here we use the numerically determined frequencies $\omega^\tau _i$ of the cosine billiard for comparison. All relative errors $\Delta_i$ are lower than $10^{-2}$.

\subsection{Phase-space representation}
\label{ch:CosineBilliardPhaseSpaceRepresentation} 

In the following we introduce more suitable coordinates $(\Q,\P)$ giving a continuous representation of trajectories in phase space. In these coordinates we determine an initial integrable approximation $\HregNod(\Q,\P)$ of $H(\Q,\P)$, which we then improve by an iterative application of canonical transformations. Finally by returning to the original coordinates $(\q,\p)$ we find the integrable approximation $\Hreg(\q,\p)$ of $H(\q,\p)$.

\subsubsection{Transformation to a system with continuous trajectories}
\label{ch:CosineBilliardRepair}

A basic feature of billiards is that the momentum $\p(t)$ is discontinuous due to reflections at the boundary. 
Therefore regular tori in phase space consist of disjoint parts which are related at the boundary by non-trivial maps, see App.~\ref{app:CosineTransformation}, Eqs.~\eqref{eq:ReflexionMomentumLowerWall} and \eqref{eq:ReflexionMomentumUpperWall}. It seems difficult to directly find an integrable approximation $\Hreg(\q,\p)$ whose trajectories reproduce these properties.
To overcome this problem, we introduce a canonical transformation
\begin{align}
 \TCal: (\q,\p) &\mapsto(\Q,\P),\label{eq:CosineBilliardTransformationToQPAbstract}
\end{align} such that in the new coordinates the regular trajectories of $H$ become continuous in phase space. 

\insertfigure{tb}{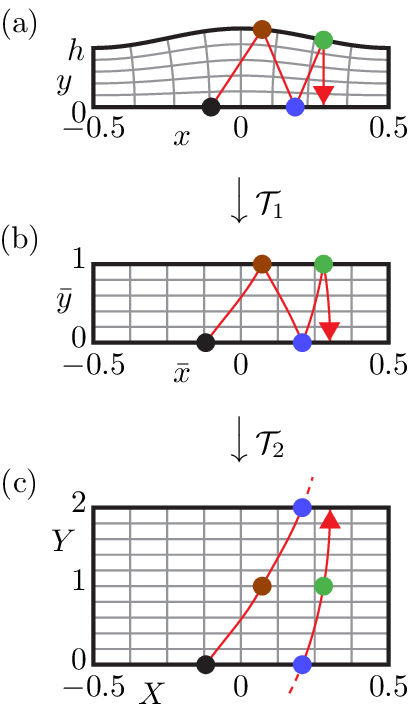}{(color online) (a) Cosine billiard and a trajectory (red line) in the coordinates (a) $\q=(x,y)$, (b) $\qbar=(\bar{x},\bar{y})$, and (c) $\Q=(X,Y)$ using the transformations $\TCal_1$, Eq.~\eqref{eq:CosineTransformationToQPStep1}, and $\TCal_2$, Eqs.~\eqref{eq:CosineTransformationToQPStep2a} to \eqref{eq:CosineTransformationToQPStep2d}. The colored dots correspond to the reflection points of the trajectory in (a). The gray grid shows lines of constant $X$ and lines of constant $Y$.}{fig:CosineBilliardTransformationToNiceSystem}
In the following we construct the transformation $\TCal$, as a composition of two transformations
\begin{align}
 \TCal &= \TCal_2\circ \TCal_1. \label{eq:CosineTransformationToQPComposition}
\end{align} Here $\TCal_1$ is given by a point transformation
\begin{align}
 \TCal_1 : &\vrr{\q}{\p}\mapsto \vrr{\qbar(\q)}{\pbar(\q,\p)}, \label{eq:CosineTransformationToQPStep1}
\end{align}
which we derive in App.~\ref{app:CosineTransformation} for a general class of billiards. As shown in Fig.~\ref{fig:CosineBilliardTransformationToNiceSystem}, this transformation maps the cosine billiard $H(\q,\p)$, see Fig.~\ref{fig:CosineBilliardTransformationToNiceSystem}(a), to a system $H(\qbar,\pbar)$ with a rectangular spatial domain, see Fig.~\ref{fig:CosineBilliardTransformationToNiceSystem}(b). This new system can be thought of as a generalized rectangular billiard, as it provides elastic reflections at the boundary but has a nontrivial time evolution. 

As the second step we get rid of momentum inversions when the particle hits the upper boundary. For this we apply an unfolding operation $\TCal_2:(\qbar,\pbar)\mapsto(\Q,\P)$, explicitly given by
\begin{align}
  X &= \bar{x} \label{eq:CosineTransformationToQPStep2a} \\
  Y &= \left\{ \begin{array}{cl}\bar{y} & \quad \bar{p}_y\geq 0 \\ 2-\bar{y} & \quad \bar{p}_y<0\end{array} \right., \label{eq:CosineTransformationToQPStep2b} \\
  P_x&= \bar{p}_x,\label{eq:CosineTransformationToQPStep2c} \\
  P_y &= |\bar{p}_y|,\label{eq:CosineTransformationToQPStep2d}
\end{align} where we use the notation $(\qbar,\pbar)=(\bar{x},\bar{y},\bar{p}_x,\bar{p}_y)$ and $(\Q,\P)=(X,Y,P_x,P_y)$. As shown in Fig.~\ref{fig:CosineBilliardTransformationToNiceSystem}(c), the trajectory segments which move in negative $\bar{y}$-direction are reflected with respect to the line $\bar{y}=1$. Furthermore, we impose periodic boundary conditions in $Y$-direction.

\insertfigure{tb}{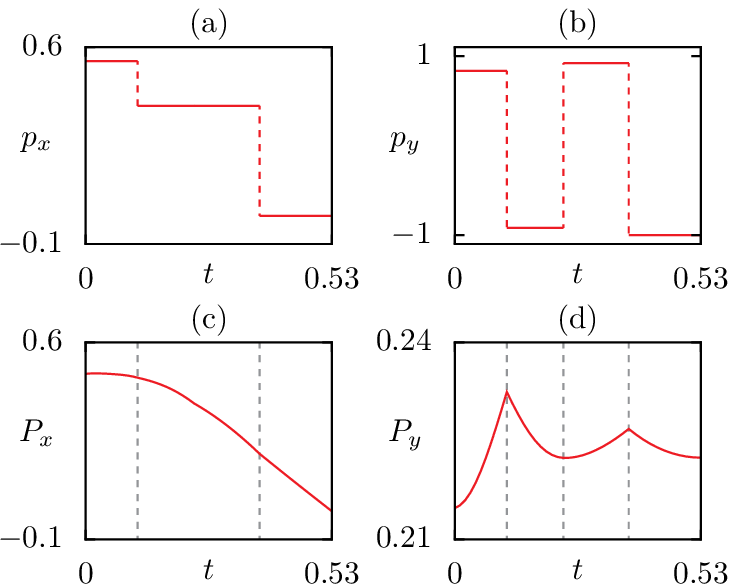}{(color online) Momentum components (a) $p_x$, (b) $p_y$ of the trajectory from Fig.~\ref{fig:CosineBilliardTransformationToNiceSystem}(a) and (c) $P_x$, (d) $P_y$ of the transformed trajectory from Fig~\ref{fig:CosineBilliardTransformationToNiceSystem}(c).}{fig:CosineBilliardTrajectories}
Due to the unfolding the trajectory components $P_x$ and $P_y$ become continuous at the positions $Y=0,1,2$ which correspond to the boundary of the original billiard, see Fig.~\ref{fig:CosineBilliardTrajectories}. This finally gives an equivalent representation $H(\Q,\P)$ of the cosine billiard where all regular trajectories are continuous in phase space. Note that the components $X$, $Y$, and $P_x$ have continuous first derivatives, while $\dot{P_y}$ has a discontinuity, see Fig.~\ref{fig:CosineBilliardTrajectories}(d). We find that this discontinuity cannot be avoided by any point transformation $\TCal_1$, Eq.~\eqref{eq:CosineTransformationToQPStep1}. Also higher order discontinuities may occur. 

Note that the transformation $\TCal$, Eq.~\eqref{eq:CosineTransformationToQPComposition}, commutes with the scaling operation~\eqref{eq:ScalingPhaseSpace} such that the Hamiltonian $H(\Q,\P)$ also fulfills the scaling relation~\eqref{eq:ScalingHamiltonian}. Moreover, both actions $\J=(J_1,J_2)$, Eq.~\eqref{eq:GeneralActionIntegrals}, are preserved under the transformation $\TCal$ such that the previously determined action representation $\HregNodCal(\J)$ remains valid for $H(\Q,\P)$.

\subsubsection{Initial integrable approximation}
\label{ch:CosineBilliardInitialIntegrableApproximation}

Starting from $\HregNodCal(\J)$, we determine the initial integrable approximation $\HregNod(\Q,\P)$ following Sec.~\ref{ch:GeneralPhaseSpaceRepresentation}. For this we choose a canonical transformation $T_0:(\bPhi,\J)\mapsto(\Q,\P)$, as in Eq.~\eqref{eq:GeneralTransformationToqp} with $(\q,\p)$ replaced by $(\Q,\P)$, given by 
\begin{align}\label{eq:TransformationIntegrableToLibRotX}
 X &= \sqrt{\frac{2 J_1}{\delta J_2}}\cos\phi_1,\\
\label{eq:TransformationIntegrableToLibRotPx}
 P_x &= -\sqrt{2\delta J_1 J_2}\sin\phi_1,\\
\label{eq:TransformationIntegrableToLibRotY}
 Y &= \frac{1}{\pi}\phi_2 + \frac{J_1}{\pi J_2}\sin\phi_1\cos\phi_1,\\
\label{eq:TransformationIntegrableToLibRotPy}
 P_y &= \pi J_2.
\end{align}
The first two equations lead to a rotation in the $(X,P_x)$-plane, with action $J_1$ and half-axis ratio $\delta J_2$, see the thin red lines in Fig.~\ref{fig:CosineBilliardCascade}(b). 
We choose the constant parameter $\delta$ to reproduce the linearized dynamics around the stable periodic orbit at $Y=0$. 
For this we derive an analytical expression for $\delta$ as a function of the cosine billiard parameters $\h$ and $\w$, see App.~\ref{app:DeltaCosine}, Eq.~\eqref{eq:TransformationParameterDelta}. 
Equations~\eqref{eq:TransformationIntegrableToLibRotY} and \eqref{eq:TransformationIntegrableToLibRotPy} describe a motion with constant momentum $P_y$, see the thin red lines in Fig.~\ref{fig:CosineBilliardCascade}(e). On average $Y$ increases linearly with $\phi_2$ such that $\phi_2\in[0,2\pi[$ is mapped to $Y\in[0,2[$ with periodic boundary conditions, see Fig.~\ref{fig:CosineBilliardTransformationToNiceSystem}(c). The second, oscillatory term in Eq.~\eqref{eq:TransformationIntegrableToLibRotY} is required to make the transformation canonical.

We finally obtain an initial integrable approximation $\HregNod(\Q,\P)=\HregNodCal[\J(\Q,\P)]$, which still obeys the scaling property, Eq.~\eqref{eq:ScalingHamiltonian}. Figures~\ref{fig:CosineBilliardCascade}(b) and (e) show that the tori of $H$ and $\HregNod$ agree close to the central trajectory. In the outer parts of the regular island $\HregNod$ needs to be improved.

\subsubsection{Family of canonical transformations}
\label{ch:CosineBilliardFamilyOfCanonicalTransformations}
\insertbigfigure{bt}{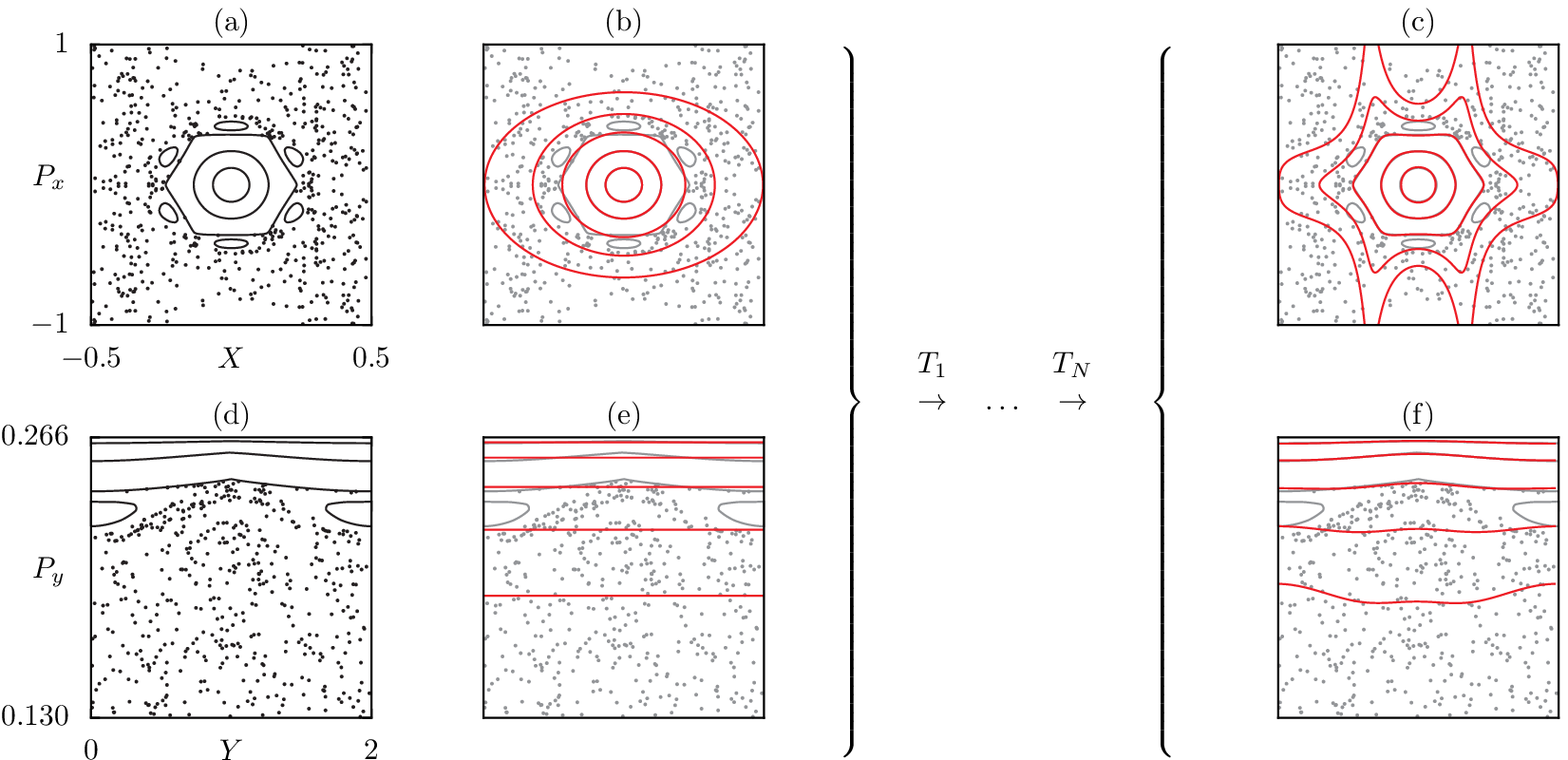}{(color online) Poincar\'e sections $\Sigma_1$ and $\Sigma_2$ for the cosine billiard $H(\Q,\P)$, (a) and (d), its initial integrable approximation $\HregNod(\Q,\P)$, (b) and (e), and its final integrable approximation $\Hreg(\Q,\P)$, (c) and (f).}{fig:CosineBilliardCascade}

We define a family of canonical transformations $\Ta$ given by the generating function~\eqref{eq:GeneralGenerator},
\begin{align}
 F^\ba(\Q,\P') &= \Q\P' +G(\Q,\P'),\label{eq:GeneratorUsingG}
\end{align} with the perturbation
\begin{align}
 G (\Q,\P') &= \sum_{\nu=1}^r a_\nu G_\nu (\Q,\P')\label{eq:GasSumOverGnu}.
\end{align} 
In the following we specify $G(\Q,\P')$ according to (i) the scaling relation, (ii) the symmetries, and (iii) the phase-space geometry.

(i) To preserve the scaling property~\eqref{eq:ScalingHamiltonian} the transformation has to commute with the scaling operation~\eqref{eq:ScalingPhaseSpace}. From Eqs.~\eqref{eq:Type21} and \eqref{eq:Type22} one gets the scaling condition 
\begin{align}
 G(\Q,\lambda \P') &= \lambda G(\Q,\P').\label{eq:ScalingGnu}
\end{align} This is fulfilled by the form 
\begin{align}
 G(\Q,\P') &= P'g(\Q,\theta'), 
\end{align} which uses the polar representation \hbox{$\P'=(P'\cos\theta',P'\sin\theta')$} with $\theta'\in[0,\pi]$ and an arbitrary function $g(X,Y,\theta')$ which depends on 3 variables only. 

(ii) The systems $H$ and $\HregNod$ both have two symmetries, namely the parity in $X$-direction
\begin{align}
 (X,Y,P_x,P_y) &\mapsto (-X, Y,-P_x,P_y), \label{eq:SymmetryQPParity}
\end{align} 
and the time reversal symmetry
\begin{align}
 (X,Y,P_x,P_y) &\mapsto (X,2- Y,-P_x,P_y), \label{eq:SymmetryQPTimeReversal}
\end{align} which follows from $(\q,\p)\mapsto (\q,-\p)$ using Eqs.~\eqref{eq:CosineTransformationToQPStep1}, \eqref{eq:CosineTransformationToQPStep2b} 
and \eqref{eq:CosineTransformationToQPStep2d}.
To ensure that the canonical transformations, Eqs.~\eqref{eq:Type21} and \eqref{eq:Type22}, commute with these symmetry transformations, the function $g$ must satisfy
\begin{align}
 g(-X,Y,\pi-\theta') &= g(X,Y,\theta'),\label{eq:Symmetrygnu1}\\
 g(X,2-Y,\pi-\theta') &= -g(X,Y,\theta').\label{eq:Symmetrygnu2}
\end{align}

(iii) To match the $Y$-periodic structure of the phase space we require
\begin{align}
 g(X,Y+2,\theta') &= g(X,Y,\theta').
\end{align}

We express $g$ as a truncated Fourier series in $X$, $Y$, and $\theta'$ with periodicities $\LCal_x=1$, $\LCal_y=2$, and $\LCal_\theta=2\pi$. This leads to the expansion
\begin{align}
 g(X,Y,\theta') &= \sum_{n=0}^{\NCal_x}\sum_{m=1}^{\NCal_y}\sum_{l=0}^{\NCal_\theta} a_{nml}^{1} f^+_n\left(\frac{X}{\LCal_x}\right)f^-_{m}\left(\frac{Y-1}{\LCal_y}\right) \nonumber \\
                & \qquad\qquad \qquad\qquad\times f^+_l\left(\frac{\theta'-\tfrac{\pi}{2}}{\LCal_\theta}\right) \nonumber \\
                & +\sum_{n=1}^{\NCal_x}\sum_{m=0}^{\NCal_y}\sum_{l=1}^{\NCal_\theta} a_{nml}^2 f^-_{n}\left(\frac{X}{\LCal_x}\right)f^+_{m}\left(\frac{Y-1}{\LCal_y}\right) \nonumber \\ 
                & \qquad\qquad \qquad\qquad\times f^-_{l}\left(\frac{\theta'-\tfrac{\pi}{2}}{\LCal_\theta}\right),  \label{eq:CosineBilliardFinalPerturbation}
\end{align} with the Fourier basis functions $f_\nu^\pm$ from Eqs.~\eqref{eq:FourierBasisPlus} and \eqref{eq:FourierBasisMinus}. The orders $\NCal_{x,y,\theta}$ can still be chosen. None of the other six combinations occurs, e.g., $f_n^+ f_m^+ f_l^+$, as they would violate the symmetry conditions~\eqref{eq:Symmetrygnu1} and \eqref{eq:Symmetrygnu2}. 

Finally, the family of canonical transformations $\Ta$ is defined by the generating function~\eqref{eq:GeneratorUsingG}. With this family the iterative improvement will be performed.

\subsubsection{Iterative Improvement}
\label{ch:CosineBilliardImplementation}

\insertfigure{b}{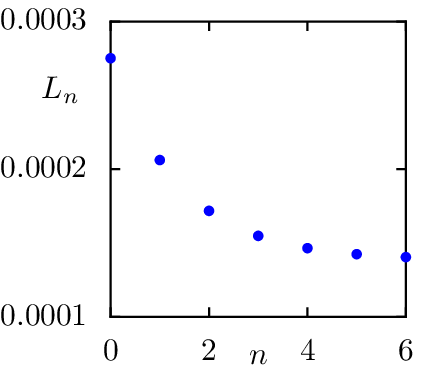}{(color online) Cost function $L$ vs. iteration step $n$.}{fig:CosineBilliardDistance}
In order to apply the iterative improvement, we evaluate the cost function $L(\ba)$, Eq.~\eqref{eq:GeneralCostFunction}, in $(\Q,\P)$-space. For this we sum over $10$ tori $\tau$ and $9000$ times $t$ for each torus. We checked that the points of the trajectories at these times are well distributed on the tori. 

We define a family of canonical transformations $\Ta$ using the generating function~\eqref{eq:GeneratorUsingG} with $\NCal_x=\NCal_y=\NCal_\theta=2$. Then the linear Eq.~\eqref{eq:GeneralLinearSystemOfEquations} is solved and the solution is rescaled using a damping factor $\eta=0.3$, Eq.~\eqref{eq:Damping}, giving the parameter $\ba_1$. This specifies the first canonical transformation $T_1=T^{\ba_1}$, which maps the tori of $\HregNod$ closer to the tori of $H$. We iteratively repeat this procedure based on the transformed tori, see Fig.~\ref{fig:CosineBilliardCascade}. After $N=6$ steps convergence is achieved. 
The final integrable approximation $\Hreg^N (\Q,\P)$ is given by Eq.~\eqref{eq:GeneralHreg1ByPluggingInTransformation}. 
In Figs.~\ref{fig:CosineBilliardCascade}(c) and (f) we compare the Poincar\'e sections of $H(\Q,\P)$ and $\Hreg^N(\Q,\P)$. 
We have confirmed that the improvement is of the same quality also in other sections of phase space (not shown). 
Fig.~\ref{fig:CosineBilliardDistance} shows the evaluated cost function $L_n$ after each iteration step $n$. 
The final value of the cost function $L_N$ is about a factor of $2$ larger than for the 2D example, see Fig.~\ref{fig:StandardMapCostFunction}. However, as $L$ is defined as a squared distance, Eq.~\eqref{eq:GeneralCostFunction}, it scales linear in the number of degrees of freedom. From that perspective the results can be seen as comparable in quality.
One would expect further improvements when choosing a larger family of canonical transformations by increasing the orders $\NCal_{x,y,\theta}$ in Eq.~\eqref{eq:CosineBilliardFinalPerturbation}.

%%%%%%%%%%%%%%%%%%%%%%%%%%%%%%%%%%%%%%%%%%%%%%%%%%%%%%%%%%%%%%%%%%%%%%%%%%%%%%%

\section{Summary and outlook}
\label{ch:Summary}

In this paper we introduce the iterative canonical transformation method which determines an integrable approximation $\Hreg$ of a Hamiltonian system $H$ with a mixed phase space. The dynamics of this integrable approximation $\Hreg$ resembles the regular motion of $H$ and extends it beyond the regular region. In contrast to other approaches the iterative canonical transformation method can be applied to strongly perturbed systems. We present its application to the standard map and to the 2D cosine billiard. For both of these generic systems we find a very good agreement between the regular dynamics of $H$ and $\Hreg$ as well as a reasonable extension of the motion beyond the regular region. 
This integrable extension is valuable quantum mechanically, as it leads to regular basis states outside the regular region, which is helpful for predicting tunneling rates \cite{BaeKetLoeSch2008,BaeKetLoe2010,LoeBaeKet2010,MerEtAl}.

The method is applicable to higher dimensions, e.g.\ for a Hamiltonian system with $f$ degrees of freedom. 
Around fully elliptic periodic orbits one has nested $f$-dimensional tori, which constitute a regular region in phase space.
A visualisation of such a higher dimensional region in the case of a 4D map is shown in Ref.~\cite{RicLanBaeKet2013}.
The application of the iterative canonical transformation method should make the prediction of tunneling rates in such higher dimensional systems possible.

%%%%%%%%%%%%%%%%%%%%%%%%%%%%%%%%%%%%%%%%%%%%%%%%%%%%%%%%%%%%%%%%%%%%%%%%%%%%%%%

\section*{Acknowledgments}

We are grateful for the financial support through the DFG Forschergruppe 760 
\textit{Scattering systems with complex dynamics}.

%%%%%%%%%%%%%%%%%%%%%%%%%%%%%%%%%%%%%%%%%%%%%%%%%%%%%%%%%%%%%%%%%%%%%%%%%%%%%%%

\appendix

\section{Algorithmic overview}
\label{app:AlgorithmicOverview}
The algorithm of the iterative canonical transformation method proceeds in the following steps:
\begin{enumerate}
  \item For a chosen set of tori $\tau$ of $H$, compute the actions $\J^\tau$ and frequencies $\bomega^\tau$. 
  \item Determine $\HregNodCal(\J)$ in action representation by minimizing Eq.~\eqref{eq:HregNodGeneralAnsatzEnergy}.
  \item Define $\HregNod(\q,\p)$ in phase-space representation by choosing a simple transformation $T_0$, Eq.~\eqref{eq:GeneralTransformationToqp}, that roughly mimics the shape of the tori of $H(\q,\p)$.
  \item For a chosen set of tori $\tau$ of $H$ determine a sample of points $\xttau$ at times $t$.
  \item For $n=0,1,\hdots,N-1:$
  \begin{enumerate}
    \item Compute the points $\xttaun$, Eqs.~\eqref{eq:ReferencePointsRecoursion} or \eqref{eq:ReferencePointsInitial}, where the transformation $T_n$ is evaluated numerically using Eqs.~\eqref{eq:Type21} and \eqref{eq:Type22}.
    \item Compute the coefficients $B_\nu$ and $C_{\mu\nu}$ of the cost function $L(\ba)$, Eqs.~\eqref{eq:CoeffBnu} and \eqref{eq:CoeffCmunu}. 
    \item Determine $\ba$ by solving Eq.~\eqref{eq:GeneralLinearSystemOfEquations} and possibly applying a damping, Eq.~\eqref{eq:Damping}.
    \item Set $T_{n+1} := \Ta$.

  \end{enumerate} 
  \item Determine $\Hreg^N$ with Eq.~\eqref{eq:GeneralHreg1ByPluggingInTransformation}.
\end{enumerate}

\section{Scaling relations}
\label{app:Scaling}
In the following we consider the class of Hamiltonians $H$ with the scaling property~\eqref{eq:ScalingHamiltonian}. 
For any solution $(\q(t),\p(t))$ of such a system, we show that
\newcommand{\qla}{\q_\lambda}
\newcommand{\pla}{\p_\lambda}
\begin{align}
 \qla(t) &:= \q(\lambda t), \label{eq:ScaledTrajq}\\
 \pla(t) &:= \lambda\p(\lambda t), \label{eq:ScaledTrajp}
\end{align} is also a solution of Hamilton's equations \eqref{eq:HamiltonianMotion}:
\begin{align}
 \dot{\qla}(t) &= \lambda \dot{\q}(\lambda t),\\
  &= \lambda \DD{H}{\p}[\q(\lambda t), \p(\lambda t)],\\
  &= \DD{H}{\p}[\q(\lambda t), \lambda \p(\lambda t)],\\
  &= \DD{H}{\p}[\qla(t), \pla(t)].
\end{align} Here we used Eqs.~\eqref{eq:HamiltonianMotion}, \eqref{eq:ScaledTrajq}, \eqref{eq:ScaledTrajp} and the scaling 
\begin{align}
  \DD{H}{\p}(\q,\lambda\p)& = \lambda \DD{H}{\p}(\q,\p).\label{eq:ScalingHDerivativesp}
\end{align} which follows from differentiating Eq.~\eqref{eq:ScalingHamiltonian} with respect to $\p$. The corresponding statement for $\dot{\pla}(t)$ is obtained analogously.

We see that to each solution $(\q(t),\p(t))$ we can associate a family of other solutions $(\qla(t),\pla(t))$ using the scaling operations
\begin{align}
 (\q,\p,t) &\mapsto (\q,\lambda \p, \lambda^{-1}t), \label{eq:ScalingPhaseSpace}\\
 E &\mapsto \lambda^2 E,\label{eq:ScalingEnergy}\\ 
 \J &\mapsto \lambda \J,\label{eq:ScalingAction}
\end{align} where the last one follows from Eq.~\eqref{eq:GeneralActionIntegrals}. Hence, in such scaling systems each solution with positive energy $E$ is related to a corresponding solution of energy $E=1$ by using a scaling factor $\lambda=1/\sqrt{E}$. Finally we point out that because of Eqs.~\eqref{eq:ScalingEnergy} and \eqref{eq:ScalingAction} the scaling property \eqref{eq:ScalingHamiltonian} for the phase-space representation $H(\q,\p)$ translates to an analogous scaling property,
\begin{align}
 \HCal(\lambda\J) &= \lambda^2 \HCal(\J),
\end{align} for the action representation of the system.

\section{Billiard transformation}
\label{app:CosineTransformation}
\insertfigure{bt}{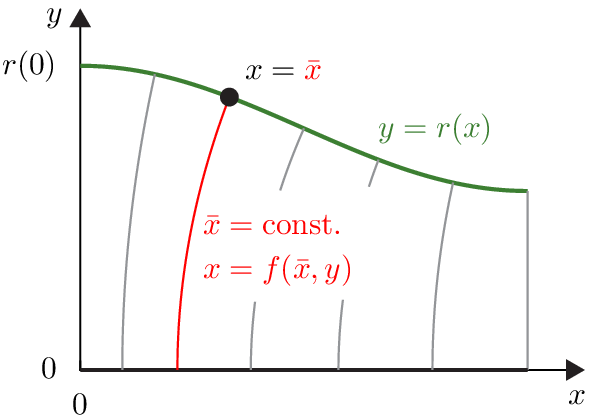}{(color online) Visualization of the point transformation $(x,y)\mapsto(\xbar,\ybar)$ for the cosine billiard. The billiard domain is filled with a family of curves (vertical lines) that represent the contour lines of $\xbar(x,y)$ and intersect the upper boundary $y=r(x)$ at $x=\xbar$.}{fig:CosineBilliardCurves}

We derive the explicit expression for the canonical transformation $\TCal_1$, Eq.~\eqref{eq:CosineTransformationToQPStep1}, which transforms the cosine billiard to a system confined in a rectangular domain with elastic reflections, see Figs.~\ref{fig:CosineBilliardTransformationToNiceSystem}(a) and (b). Here we consider the more general class of billiards given by a rectangle whose upper boundary is replaced by a curve $y=r(x)$. First we introduce a point transformation $(x,y)\mapsto(\xbar,\ybar)$ in position space. We fill the inner billiard domain with a family of curves (see the vertical curves in Fig.~\ref{fig:CosineBilliardCurves}) which define the contour lines of $\xbar(x,y)$. This family of curves is given by $x=f(\xbar,y)$ where the function $f$ has to be determined. The transformation equations are
\begin{align}
 x(\xbar,\ybar) &= f[\xbar,y(\xbar,\ybar)],\label{eq:ExplicitTransformationx}\\
 y(\xbar,\ybar) &= r(\xbar)\ybar. \label{eq:ExplicitTransformationy}
\end{align} For practical reasons we require the coordinates $x$ and $\xbar$ to coincide at the upper boundary which leads to
\begin{align}
 f[\xbar,r(\xbar)] &= \xbar.\label{eq:RWPForf1}
\end{align} and ensures that $\ybar\in[0,1]$. This point transformation implies a canonical transformation in phase space, described by the generating function
\begin{align}
 F(\xbar,\ybar,p_x,p_y) &= x(\xbar,\ybar)p_x+y(\xbar,\ybar)p_y. 
\end{align} The corresponding momentum transformation is
\begin{align}
 \vrr{p_x(\xbar,\ybar,\pxbar,\pybar)}{p_y(\xbar,\ybar,\pxbar,\pybar)} &= \left.\vrr{\DD{x}{\xbar} & \DD{y}{\xbar}}{\DD{x}{\ybar} & \DD{y}{\ybar}}\right|_{(\xbar,\ybar)}^{-1}\vrr{\pxbar}{\pybar}.\label{eq:ExplicitTransformationp}
\end{align} 
Equations~\eqref{eq:ExplicitTransformationx}, \eqref{eq:ExplicitTransformationy}, and \eqref{eq:ExplicitTransformationp} define a canonical transformation $\TCal_1:(x,y,p_x,p_y)\mapsto(\xbar,\ybar,\pxbar,\pybar)$ which depends on the function $f$. We specify $f$ by requiring certain properties for the reflections at the boundary of the transformed system. 

In the old coordinates the reflections are given by the operators 
\begin{align}
 R_1: &\quad \vrr{p_x}{p_y}\mapsto\vrr{p_x}{-p_y}, \label{eq:ReflexionMomentumLowerWall}\\
 R_2: &\quad \vrr{p_x}{p_y}\mapsto \vrr{p_x}{-p_y} + \frac{2 r'(x)}{1+r'(x)^2}
\vrr{p_y-r'(x)p_x}{p_x+r'(x)p_y}, \label{eq:ReflexionMomentumUpperWall}
\end{align} at the lower ($y=0$) and the upper boundary ($y=r(x)$), respectively. 
We require the transformed reflections \hbox{$\TCal_1\circ R_{1,2}\circ  \TCal_1^{-1}$} to take the simple form
\begin{align}
 \vrr{\pxbar}{\pybar} & \mapsto\vrr{\pxbar}{-\pybar}
\end{align} for $R_1$ at the lower boundary ($\ybar=0$) and for $R_2$ at the upper boundary ($\ybar=1$). This leads to the following conditions for the function $f$
\begin{align}
 \DD{f}{\xbar}[\xbar,r(\xbar)] &= 1+r'(\xbar)^2,\label{eq:RWPForf2}\\
 \DD{f}{y}[\xbar,r(\xbar)] &= -r'(\xbar),\label{eq:RWPForf3}\\
 \DD{f}{y}(\xbar,0) &= 0. \label{eq:RWPForf4}
\end{align} 
We solve the set of conditions \eqref{eq:RWPForf1}, \eqref{eq:RWPForf2}, \eqref{eq:RWPForf3}, and \eqref{eq:RWPForf4} and obtain
\begin{align}
 f(\xbar,y)  &= \xbar+\tfrac{1}{2}r(\xbar)r'(\xbar)-\frac{r'(\xbar)y^2}{2r(\xbar)}.
\end{align}
We verify that for the special case of the cosine billiard with $r(x)$ given by Eq.~\eqref{eq:CosineBilliardBoundary}, this canonical transformation is invertible everywhere in phase space if the parameters satisfy
\begin{align}
 \pi^2 \w(\h+\w) &<1.
\end{align} This includes all configurations of interest for which the central orbit is stable.

\section{Derivation of the parameter $\delta$}
\label{app:DeltaCosine}

We derive the parameter $\delta$ relevant for the half-axis ratio of the regular tori, see Eqs.~\eqref{eq:TransformationIntegrableToLibRotX} and \eqref{eq:TransformationIntegrableToLibRotPx}. First we consider the Poincar\'e map $M:(x,p_x)\mapsto(x',p_x')$ on the section $\Sigma_1$. This map has a fixed point $(x^*,p_x ^*)=(0,0)$ which corresponds to the central, stable periodic orbit of the billiard. 
Using Eq.~\eqref{eq:GeneralActionIntegrals} we obtain its action $J_2$ as
\begin{align}
 J_2 &= \frac{\ell}{2\pi},\label{eq:BilliardSOSMapJ2}
\end{align} where $\ell=2(\h+\w)$ is its length.

According to Ref.~\cite{SieSte1990} we compute its monodromy matrix as
\begin{align}
 \A &= \vrr{1-\ell\kappa & \ell(1-\frac{\ell\kappa}{2})}{-2\kappa & 1-\ell\kappa}. \label{eq:BilliardSOSMapMonodromyxpx}
\end{align} Here $\kappa=r''(0)=2\pi^2\w$ is the curvature of the upper boundary at the reflection point of the stable periodic orbit.
The half-axis ratio $\sigma$ of the local elliptic dynamics can be expressed in terms of $\A$ using Eqs.~\eqref{eq:SigmaFromMonodromyMatrix} and \eqref{eq:SigmaConstantc}.

The next step is to calculate the new half-axis ratio $\sigma'$ after the transformation~\eqref{eq:CosineBilliardTransformationToQPAbstract}. Using Eqs.~\eqref{eq:ExplicitTransformationx} and \eqref{eq:ExplicitTransformationp} we find
\begin{align}
 \sigma' &= \sigma \left(1-\frac{\ell\kappa}{4}\right)^2.\label{eq:BilliardSOSMapSigmaXPx}
\end{align} On the other hand the stable periodic orbit of $\HregNod$ has the local half-axis ratio
\begin{align}
 \sigma' &= \delta J_2,\label{eq:BilliardSOSMapSigmaJ2}
\end{align} as follows from Eqs.~\eqref{eq:TransformationIntegrableToLibRotX} and \eqref{eq:TransformationIntegrableToLibRotPx}. Finally, by combining Eqs.~\eqref{eq:BilliardSOSMapSigmaXPx}, \eqref{eq:BilliardSOSMapSigmaJ2}, and \eqref{eq:BilliardSOSMapJ2} we obtain
\newcommand{\termmodplus}{\left|2\kappa+\ell(1-\frac{\ell\kappa}{2})\right|}
\newcommand{\termmodminus}{\left|2\kappa-\ell(1-\frac{\ell\kappa}{2})\right|}
\begin{align}
 \delta &= \frac{2\pi}{\ell}\sqrt{\frac{\termmodplus-\termmodminus}{\termmodplus+\termmodminus}}\left(1-\frac{\ell\kappa}{4}\right)^2.\label{eq:TransformationParameterDelta}
\end{align} \quad

\end{document}